# Nested Ordered Sets and their Use for Data Modelling

Alexandr Savinov

Dept. of Computer Science III, University of Bonn,
Roemerstr. 164, 53117 Bonn, Germany

Institute of Mathematics and Computer Science, Academy of Sciences of Moldova
str. Academiei 5, 2028 Chisinau, Moldova

savinov@iai.uni-bonn.de
http://conceptoriented.org/savinov

**Abstract.** In this paper we present a new approach to data modelling, called the concept-oriented model (CoM), and describe its main features and characteristics including data semantics and operations. The distinguishing feature of this model is that it is based on the formalism of nested ordered sets where any element participates in two structures simultaneously: hierarchical (nested) and multi-dimensional (ordered). An element of the model is postulated to consist of two parts, called identity and entity, and the whole approach can be naturally broken into two branches: identity modelling and entity modelling. We also propose a new query language with the main construct, called concept, defined as a pair of two classes: identity class and entity class. We describe how its operations of projection, de-projection and product can be used to solve typical data modelling tasks.

# 1 Introduction

In wide sense, one of the main goals of data modelling consists in providing means for *data organization* by grouping data elements using some structure which makes data access and manipulation easier. In this paper we propose a new approach to data organization, called the concept-oriented model (CoM) which is based on a new formalism of nested ordered sets. Parts of this approach were described earlier [Sav05a, Sav05b, Sav06a, Sav06b, Sav07b, Sav09a] and in this paper we try to formalize it by changing its mechanisms if necessary. We also describe main elements of a new query language, called the concept-oriented query language, which reflects specific features of this data model.

Informally, CoM proceeds from the observation that data world has two sides or flavours postulated as the principle of duality. On one hand, data elements (of any kind and independent of their semantics) need to *exist* and for that purpose they have to be somehow *represented*. The part of an element which manifests its existence and which is responsible for its representation is referred to as *identity*. Identities are supposed to exist within their own structure having a hierarchical form similar to conventional postal addresses. On the other hand, data elements are supposed to have *semantics* which is independent of their identity and this (dual) part of an element is referred to as *entity*. What is important is that semantically elements exist within the dual structure having a multi-dimensional form where each element has a number of super-elements and sub-element. Within this structure, the meaning of an element depends on its position among other elements.

These two structures responsible for data representation and data semantics lead to separation of two branches or directions, called *identity modelling* and *entity modelling*. Yet, the main point of the proposed approach is that we cannot model these sides of the data worlds separately because they exist in inseparable unity. In particular, identity and entity are two sides of one whole like real and imaginary parts of complex numbers. Identity is passed and stored only by-value, i.e., it is a transient part which cannot be shared and exists only now and here without its own representative. Entity part of an element is a thing-in-itself which is not observable in its original form and is radically unknowable. It is passed and stored indirectly by-reference using its identity.

To describe the phenomenon of duality we propose to use the formalism of *nested ordered sets*. It is a generalization of conventional ordered sets where members of the set can themselves be ordered sets (with some constraints connecting nesting and ordering). Any element of such a set participates in two structures simultaneously and it is precisely what we need to formally substantiate the duality. What is also new here is that we propose to use *ordering relation* to describe data semantics (independent of nesting). Two operations defined on the ordered structure, called *projection* and *de-projection*, provide simple and flexible means for data access using this structure. To demonstrate how this model can be used to solve typical problems we describe a simple query language. Its main construct is called *concept* which is defined as a pair of two classes: one *identity class* and one *entity class*.

In Section 2 we provide main definitions and introduce the formalism of nested ordered sets. Section 3 describes what we mean by data semantics, its canonical representation and operations with data. In Section 4 we define main elements of the concept-oriented query language and show how typical data modelling tasks can be solved. Section 5 is a very short overview of related works and Section 6 makes concluding remarks.

# 2 Nested Ordered Sets

*Partial order* is a binary relation '$\leq$' (less than or equal to) on elements of the set $O = \{a, b, c, \ldots\}$ satisfying the properties of

- *reflexivity*, $\forall a \in O$, $a \leq a$,
- *antisymmetry*, $\forall a, b \in O$, $(a \leq b) \wedge (b \leq a) \Leftrightarrow a = b$, and
- *transitivity*, $\forall a, b, c \in O$, $(a \leq b) \wedge (b \leq c) \Rightarrow a \leq c$.

A *partially ordered set* is a set with partial order established on its elements (cf. [Dav02]). If additionally any two elements $a, b \in O$ have both a least upper bound $\sup(a,b)$ (supremum) and a greatest lower bound $\inf(a,b)$ (infimum) then it is referred to as a *lattice*. A lattice has two special elements. The greatest element $g$, called *top*, is greater than or equal to any other element of the set: $\forall a \in O \; a \leq g$. The least element $l$, called *bottom*, is less than or equal to any other element of the set: $\forall a \in O \; l \leq a$.

For data modelling purposes it is more convenient to use strict inequality '<' (less than) between elements. Another modification is that we use labels to identify elements of the ordering relation and such a structure is referred to as *labelled ordered set*. As a consequence two elements can be ordered more than once using distinct labels specified as lower index of the relation: $a <_x b$ and $a <_y b$. The transitivity condition in the presence of labels is written as follows: if $(a <_x b) \wedge (b <_y c)$ then $a <_z c$ where $z = x.y$ is a concatenation of two labels $x$ and $y$.

Labelled ordered sets can be represented by a directed acyclic graph where nodes are elements and edges are elements of the ordering relation drawn as labelled arrows from a smaller element to the greater element. We will always position a greater element higher than smaller elements. For example, if $a <_x b$ then $a$ is drawn under $b$ and arrow $x$ leads upward from node $a$ to node $b$. In the presence of top and bottom elements all paths start from bottom and finish at top. Below we define several terms which are used in CoM.

> **Definition 1 [Super- and sub-elements].** If $a <_x b$ then $a$ is a *sub-element* for $b$, and $b$ is a *super-element* for $a$.

In an ordered set graph, super-elements are positioned over this element and sub-elements are positioned under it while any arrow always leads from a sub-element to a super-element.

> **Definition 2 [Primitive elements].** Direct sub-elements of top element are called *primitive elements*.

For simplicity, we will assume that any primitive element has only one dimension connecting it with top element.

> **Definition 3 [Simple dimension and sub-dimension].** If $a <_x b$ then $x$ is referred to as a *simple dimension* (or local dimension) of the source element $a$. Dimension with the opposite direction is called *sub-dimension*.

Dimensions lead up from an element to its super-elements while sub-dimensions lead down from an element to its sub-elements.

> **Definition 4 [Source and target].** If $a <_x b$ then element $a$ is called *source* and is denoted as $\text{Sub}(x) = a$ while element $b$ is called *target* of the dimension $x$ and is denoted as $\text{Super}(x) = b$.

We will also denote the target (domain) by the dimension itself, i.e., symbol $x$ can denote both the dimension itself and (for simplicity) the element $b$ where it ends.

> **Definition 5 [Arity and cardinality].** The number of dimensions of an element is called its *arity* (dimensionality or intension): $\text{Dim}(a) = |\{x_i \,|\, \text{Sub}(x_i) = a\}|$. The number of dimensions that end in this element is referred to as its *cardinality* (extension) $\text{Card}(a) = |\{x_i \,|\, \text{Super}(x_i) = a\}|$.

Dimensionality is the number of upward arrows starting from this element and leading to its super-elements. Notice that many dimensions can lead to one super-element and this is precisely why we introduced labels. Cardinality is the number of incoming arrows while dimensionality is the number of outgoing arrows. Notice again that a sub-element can be counted many times if there are many dimensions leading from it to this element.

> **Definition 6 [Dimension].** A (complex) *dimension* $x_1.x_2.\cdots.x_k$ of element $a$ is a sequence of local dimensions (separated by dots) where first dimension belongs to $a$, $\text{Sub}(x_1) = a$, and each next dimension starts where the previous dimension ends, $\text{Sub}(x_i) = \text{Super}(x_{i-1})$, $i = 2, 3, \ldots, k$. The number $k$ of constituents in a complex dimension is referred to as *rank*.

A complex dimension is an upward path in the ordered set graph consisting of $k$ edges and leading from this element to some its direct or indirect super-element. A complex sub-dimension is a downward path from this element to some its direct or indirect sub-element.

The number of dimensions from this element to top element is referred to as a *primitive arity* (or dimensionality) of this element. We will always assume that there is only one dimension from any primitive element to top element and then primitive arity is equal to the number of dimensions from this element to primitive elements. The number of sub-dimensions to bottom element is referred to as a *primitive cardinality*. A *full arity* of an element is the number of dimensions to *all* its direct or indirect super-elements. A *full cardinality* is the number of sub-dimensions to *all* its direct or indirect sub-elements.

The same characteristics for the whole labelled lattice are those defined for top and bottom. This means that the lattice primitive/full arity is that of bottom element. And the lattice primitive/full cardinality is that of top element. In graph terms, the lattice dimensionality is the number of paths leading from bottom to top and cardinality is the number of paths from top to bottom (these numbers are apparently equal). For example, set $O$ in Fig. 1a has 3 primitive elements $d$, $e$ and $f$; it has dimensionality 5 and element $e$ has 2 sub-dimensions and one super-dimension.

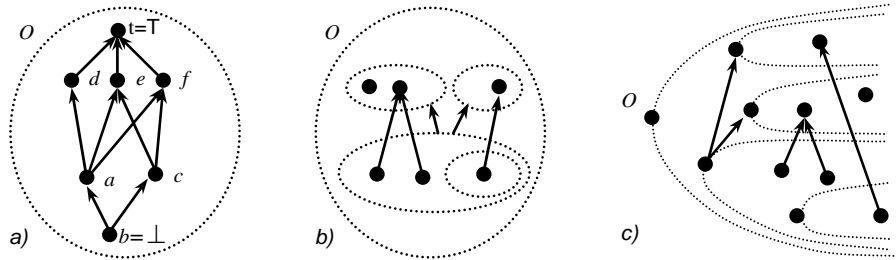

**Fig. 1.** Ordered set (a) and nested ordered set (b) and (c).

How an ordered set can be *represented*? CoM makes an assumption that sets are *inherently* ordered which means that elements have an intrinsic possibility to be ordered by containing the order in their definition. More specifically, we assumed that an element is defined as a combination of its super-elements: $e = \langle e_1, e_2, \ldots \rangle \in O$. In other words, any element by definition exists within an ordered structure because the only way to define it consists in specifying its super-elements. Elements are defined via other elements and such a structure is then interpreted as an ordered set.

One of the main assumptions of CoM, distinguishing it from other models and approaches is that ordering relation can be used to represent data semantics, i.e., data is represented as an ordered set and the meaning of each element depends on its relative position to other elements. There exist many possible interpretations of the ordering relation in the context of data modelling but we below we describe only four of them:

1. general-specific,
2. conjunction-disjunction (logical),
3. collections-combinations (grouping),
4. object-attribute-value (characterization).

The first interpretation means that a super-element is *more general* than its sub-elements and vice versa a sub-element is *more specific* than its super-elements. Arrows in the ordered set graph lead from more specific up to more general elements.

For example, if circle is known to be more specific than figure then it is represented by an upward arrow leading from the circle to a figure as one of its super-elements.

Logically, an element is equal to a *conjunction* of its super-elements and a *disjunction* of its sub-elements:

$\forall a,b,c \in O$ if $a < b$ and $a < c$ then $a = b \wedge c$ - conjunction

$\forall a,b,c \in O$ if $a > b$ and $a > c$ then $a = b \vee c$ - disjunction

In the ordered set graph this interpretation can be easily represented if all outgoing arrows are thought of as connected by conjunction while all incoming arrows are connected by disjunction.

From the point of view of grouping, an element is a *combination* of its super-elements which is interpreted as an object, record or tuple combining values in its fields. Dually, an element is a *collection* or set consisting of its sub-elements (grouping algebra [Li96] also uses this interpretation):

$\forall a,b,c \in O$ if $a < b$ and $a < c$ then $a = \langle b,c \rangle$ - combination (object, record)

$\forall a,b,c \in O$ if $a > b$ and $a > c$ then $a = \{b,c\}$ - collection (set)

For example, if we know that a department consists of employees then the employees are sub-elements of the department. On the other hand, if a person is a combination of his address and project then they are super-elements for the person.

An ordered set can be interpreted in terms of object characteristics using the assumption that super-elements are *values* characterizing this element while sub-elements are *objects* characterized by this element as a value:

$\forall a,b \in O$ if $a < b$ then $b$ is a value characterizing object $a$

For example, if an employee (sub-element) is assigned a department (super-element) then employee is a characterized object while department is a characterizing value. For this interpretation we can use dotted notation to represent ordering relation:

$\forall a,b \in O \quad a <_x b \quad \Leftrightarrow \quad a.x = b$ ( $x = b$ for object $a$)

It is important to understand that formally the model is represented by a nested ordered set while these and other possible interpretations are used only for design purposes.

We can mix these interpretations and derive that a value of an attribute is always more general than the object it characterized. Or a value is a collection consisting of objects it characterizes. For example, if bank account is a sub-element and person (account owner) is a super-element then we known that (i) owner is more general than account, (ii) owner is a disjunction of its accounts, (iii) owner is a set of its accounts, and (iv) account as an object is characterized by its owner as a value.

Defining elements (semantically) via super-elements constitutes only one half of the model structure. The other half and important distinguishing feature of CoM is that elements can themselves be ordered sets consisting of their own elements so that in addition to the ordered structure we get a nested structure. Here we use term *parent* to refer to the set where this element exists and term *child* to refer to elements this set consists of. For example, a bank consists of accounts which in turn consist of savings accounts. It is important that each child element is identified in the context of its parent element. The most interesting property is that if we combine the nested structure and the ordered structure then each element is defined dually as

- a combination of super-elements $e = \langle e_1, e_2, \ldots \rangle$, and
- a set of child elements $e = \{c_1, c_2, \ldots\}$.

Using this observation we can define what we mean by nested ordered sets.

> **Definition 7 [Nested ordered set].** A *nested ordered set* is a set where an element is defined as consisting of two parts, $e = \{c_1, c_2, \ldots\}\langle e_1, e_2, \ldots \rangle$, where the first part is a *set* of child elements, $e = \{c_1, c_2, \ldots\}$, and the second part is a *combination* of super-elements, $e = \langle e_1, e_2, \ldots \rangle$ satisfying the following conditions:
> - [hierarchical] an element can belong to only one parent set except for one root element which has no parent,

- [multi-dimensional] the structure of combinations is an ordered set,
- [syntactic constraint] if $e = \langle e_1, e_2, \ldots, e_n \rangle \in C = \langle C_1, C_2, \ldots, C_n \rangle$ then $e_i \in C_i$ (directly or indirectly), $i = 1, 2, \ldots, n$.

Syntactic constraint connects two structures so that an element may have super-elements only as children of its parent super-elements. Syntactic constraint is also important because parent dimensions are effectively "inherited" by child elements, i.e., if we order two parent elements $C_1$ and $C_2$ using dimension $x$, $C_1 <_x C_2$, then their children also can use this dimension to establish an order.

The main idea of CoM is that nested ordered sets can be used for the purposes of data modelling, i.e., to describe data organization, semantics, operations, data access, constraints and other properties and mechanisms. The main purpose of the hierarchical structure (nested sets) consists in describing how data is *represented*. It is the main tool that serves for *identity modelling*. Elements cannot change their position because this hierarchy is intended for describing identity of elements which cannot change in time. This structure is somewhat analogical to the hierarchical model of data. Identity modelling also in great extent overlaps with the concept-oriented programming. In particular, an element in the nested structure of sets is identified using a *complex identity* which is sequence of local identities where each next identity belongs to the set of the previous identity. For example, a savings account element is identified as a sequence of its bank id followed by main account id and finally ending with the savings account number. Fig. 1 b and c provide an example of a nested ordered where vertical upward arrows denote ordering relation while hierarchical structure is shown as nested elements in Fig. 1b and as a tree in Fig. 1c. Syntactic constraint means that an arrow is only possible if only the parents are also connected with such an arrow.

The main purpose of multi-dimensional structure (ordered sets) consists in describing data semantics. It is the main tool that serves for *entity modelling*. This structure can change in time by reflecting changes in the problem domain.

# 3 Data Semantics and Operations

## 3.1 Canonical Representation

Initially, data semantics in CoM is defined by the ordering relation, i.e., semantics of each concrete element is defined by its position among other elements within the ordered set. The meaning of each concrete element is defined by its super-elements, i.e., elements define their semantics via themselves. However, the problem with such a representation is that elements use their own dimensions and it is difficult to learn how they are related. This problem can be solved if we introduce a *canonical* representation where elements use one common set of dimensions. In this case all elements can be represented and manipulated as points of one space with the structure induced by the original ordered set.

In order to introduce such a canonical representation we need (i) to convert original dimensions into the set of canonical dimensions (syntax), and then (ii) to map the original elements to this new space (semantics). The result of such a transformation is a table with columns corresponding to dimensions of the canonical representation and rows corresponding to canonical semantics. The approach described in this paper consists in choosing *all* primitive dimensions of the model as the target space syntax so that one column of the target table corresponds to one path from bottom to top in the ordered set graph.

> **Definition 8 [Primitive syntax].** *Primitive syntax* is a set of all primitive dimensions $p_1, p_2, \ldots, p_N$ of the ordered set, where $\mathrm{Sub}(p_i) = B$, $\mathrm{Super}(p_i) = T$, $T$ is top element and $B$ is bottom element.

Primitive semantics is a set of rows in the primitive table where each column takes only primitive value. To define primitive semantics we need to describe rules for converting original data elements into the primitive syntax. In other words, each original data element defined as a combination of its super-elements needs to be somehow converted into one or more rows of the table with primitive syntax.

Here we describe the approach where one element produces as many target rows as it has primitive sub-dimensions. In this case the number of elements in the primitive representation is larger than the number of elements in the original multi-dimensional representation. This approach reflects the fact that one element may have many uses which are defined by its sub-elements. For example, one address could be used as an address of a person and bank. Here one address has two uses defined by two sub-dimensions and it produces two rows in the primitive semantics where one row relates to a person address and the second row relates to a bank address. Such a conversion is performed using the following rule.

> **Definition 9 [Primitive semantics].** Data element $e$ with $n$ primitive super-dimensions $d_1, d_2, \ldots, d_n$ and $m$ primitive sub-dimensions $f_1, f_2, \ldots, f_m$ produces $m$ primitive elements $r_1, r_2, \ldots, r_m$ where $r_i = \langle f_i.d_1, f_i.d_2, \ldots, f_i.d_n \rangle$, $i = 1, 2, \ldots, m$.

Here $f_i.d_j$ is some primitive dimension defined as concatenation of one primitive sub-dimension and one primitive super-dimension of this element. According to this definition each row will be a point of the hyper-cube $\Omega = P_1 \times P_2 \times \ldots \times P_N$ where $P_i = \{0,1\}$, $i = 1, 2, \ldots, N$. Points produced by one element use different primitive dimensions, i.e., they are incomparable in the space of the primitive elements just because they have different roles (like person and bank addresses). Logically, one source element produces a disjunction of sub-dimensions $r_1 \vee r_2 \vee \ldots \vee r_m$ where each constituent is a conjunction of super-dimensions $r_i = f_i.d_1 \wedge f_i.d_2 \wedge \ldots \wedge f_i.d_n$.

For example, the ordered set in Fig. 2 has 4 primitive dimensions which are mapped to the columns. Elements $e$ and $d$ produce two points of the table because they have one super-dimension and two sub-dimensions. Elements $a$ and $c$ have only one sub-dimension and produce one row each. Top element is also an empty row while bottom has all primitive dimensions.

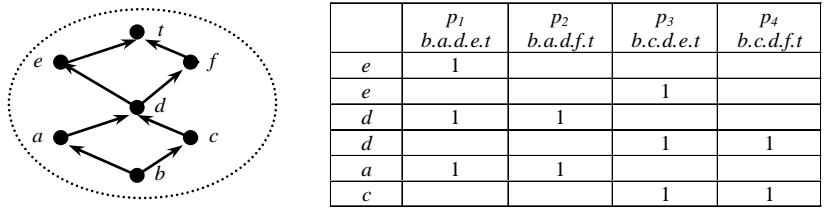

|   | $p_1$ $b.a.d.e.t$ | $p_2$ $b.a.d.f.t$ | $p_3$ $b.c.d.e.t$ | $p_4$ $b.c.d.f.t$ |
|---|---|---|---|---|
| $e$ | 1 |   |   |   |
| $e$ |   |   | 1 |   |
| $d$ | 1 | 1 |   |   |
| $d$ |   |   | 1 | 1 |
| $a$ | 1 | 1 |   |   |
| $c$ |   |   | 1 | 1 |

**Fig. 2.** Primitive semantics of data.

### 3.2 Operations

The ordered structure is very convenient for navigating over related elements in the model using super-dimensions and sub-dimensions. For this purpose we introduce two operations: projection and de-projection. Projection operation allows us to navigate upwards in the direction of more general elements.

> **Definition 10 [Simple projection].** Projection of set $E$ along simple dimension $d$ to set $D$ is a set of all their super-elements along this dimension belonging to set $D$:
> $E \rightarrow d \rightarrow D = \{e.d \mid e \in E, e.d \in D\}$.

Simple (or local) projection allow us to move one step up, i.e., to access direct super-elements. Notice that an element can be included into projection only once even if many source elements have one and the same super-element. For example, if a set of persons is projected to all addresses then we get their addresses. If many persons have one address then this address is returned only one time. For complex dimensions projection is defined recursively as follows:

> **Definition 11 [Projection].** *Projection* of set $E$ along complex dimension $d = d^1.\cdots.d^k$ of rank $k$ to intermediate sets $D^1.\cdots.D^k$, where $D^{i-1} <_{d^i} D^i$, $i = 2, \ldots, k$:
> $E \rightarrow d \rightarrow D^k = (E \rightarrow d^1.\cdots.d^{k-1} \rightarrow D^{k-1}) \rightarrow d^k \rightarrow D^k$ or
> $E \rightarrow d \rightarrow D^k = (E \rightarrow d^1 \rightarrow D^1) \rightarrow d^2.\cdots.d^k \rightarrow D^k$.

For example, if products are characterized by a sub-category (say, cars) which in turn has a main category (say, vehicles) then we can project a set of products items along these two segments and get a set of all main categories for these products.

Projection has a natural geometrical interpretation if a collection is viewed as a multi-dimensional space with super-collections as axes and super-elements as coordinates along these axes. The important specific feature of CoM is that the axes are defined hierarchically because coordinates (super-elements) may have their own coordinates and so on up to the primitive elements (see [Sav05a] section for more discussion of this interpretation).

The opposite operation, called de-projection, makes it possible to move downwards in the direction of more specific elements.

> **Definition 12 [De-projection].** *De-projection* of set $E$ along sub-dimension $f$ to set $F$ is a set of all sub-elements from $F$ along this sub-dimension: $e \leftarrow f \leftarrow F = \{s \in F \mid s.f = e\}$.

In other words, de-projection of a set consists of all elements which are projected to this set along the specified dimension. Notice that in contrast to projection, elements cannot be de-projected to the same sub-element and hence de-projection always consists of unique elements. For example, we can de-project a set of addresses to persons and get all persons with these addresses. Projection and de-projection operations can be easily defined on primitive representation using coverage relation [Sav07b] between its rows.

Projection and de-projection operations allow us to navigate over multi-dimensional structure by moving up and down to super- or sub-elements. According to the interpretation of the ordering relation we will obtain more abstract or more specific elements, i.e., by applying these operations the current level of details can be changed. However, these two operations do not allow us to navigate over hierarchical structure and hence some parts of the model can be unreachable. In order to overcome this problem we can use a special dimension, called 'parent', to refer to the parent element in the nested structure. Such operations using dimension 'parent' can be called hierarchical projection and de-projection (while these operations with normal dimensions are called multi-dimensional projection and de-projection). Hierarchical projection returns a set of all parents for this set of elements. Hierarchical de-projection returns all child elements for this set. For example, if one bank is a set consisting of its accounts then we can get a set of banks for given accounts via projection. And vice versa, given a set of banks we can get all their accounts via de-projection.

CoM also supports more traditional operations such as product, intersection and union of sets of elements. A product of two sets is a new set where each element has a new identity and combines elements of the source sets as its super-elements: $C = D_1 \times D_2 \times \ldots \times D_n = \{\langle e_1, e_2, \ldots, e_n \rangle, e_i \in D_i\}$. So it is the Cartesian product with the only difference that elements of the product are treated as sub-elements which combine the source super-elements.

# 4 Concept-Oriented Query Language

## 4.1 Concept and Concept Schema

Both identity and entity have their own structure and behaviour and hence they can be described by classes which are called *identity class* and *entity class*, respectively. In order to describe them in their inseparable unity we propose a new data modelling construct, called *concept*, which is defined as a pair of two classes – identity class and entity class. For example, if banks are identified by their IBAN number while address and name are their entity properties then the type of such elements can be specified via the following concept:

```
CONCEPT Bank                              // Concept name
  IDENTITY                                // Identity class
    CHAR(16) iban                         // Identity class field
  ENTITY                                  // Entity class
    CHAR(64) name,                        // Entity class fields
    Address address                       // Identity of address
```

Concepts are used instead of conventional classes as types when instantiating either (i) individual elements or (ii) their collections. When concept is used as a type of a variable (field, parameter or return value) then this variable will contain only identity of the element. For example, entity field `address` in concept `Bank` defined above will contain identity of concept `Address` which is its type.

For collections we need actually two types: one for the collection itself (like `Table` or `Array`) and the other for its elements (like `Bank` or `Address`). The collection type is normally provided by DBMS while the type of its elements has to be provided by the user as some concept name. For example, a table for storing bank records is created as follows:

```
CREATE TABLE Banks CONCEPT Bank
```

The new collection `Banks` will contain only entities of concept `Bank` while identities will play a role of references providing indirect access to the entities.

To model hierarchical structure CoM uses concept *inclusion relation*. Just as concepts generalize conventional classes, concept inclusion generalizes class inheritance [Sav07a]. If a new concept has a parent concept then it is specified after keyword 'IN'. For example (Fig. 3a), bank accounts exist in the context of their bank and hence concept `Account` has to be included in concept `Bank`:

```
CONCEPT Account IN Bank                              // Parent concept
   IDENTITY CHAR(8) accNo
   ENTITY DOUBLE balance, Person owner
```

The main consequence of such a declaration is that bank accounts will be identified by two numbers: account number and the parent bank identity where this account has been created.

>**Definition 15 [Complex identity].** *Complex identity* is a sequence of identities, called *identity segments*, where each previous segment has a type of the parent concept.

Complex identities are used to represent elements within a hierarchy where each identity segment is a local identifier in the context of its parent element. If concept `SavingsAccount` were included in concept `Account` then its elements would be identified by three segments (bank, account, sub-account). Just as individual segments, complex identities are passed by-value and are used to represent entities. For example, a variable or field having a type of `Account` will store complex identity consisting of bank id and account id. In many cases we might need to store only local identity (without some parent segments) and this can be done using operations for reference length control [Sav07a].

>**Definition 16 [Complex entity].** *Complex entity* is a sequence of entities, called *entity segments*, represented by segments of a complex identity.

Segments of a complex entity need not to reside together side-by-side because they have their individual references. For example, account entity and bank entity can be stored in different tables.

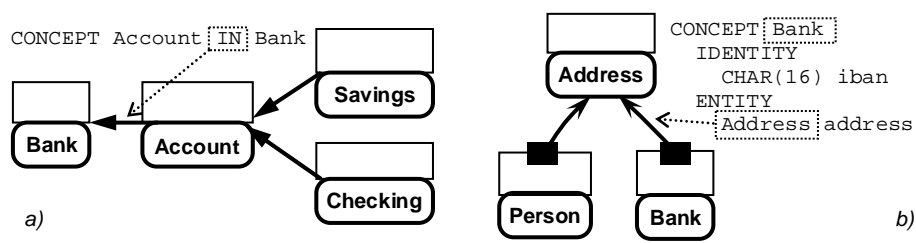

**Fig. 3.** Concept inclusion (a) and concept ordering (b).

The major difference from inheritance is that instances of concepts exist within a hierarchy while instances of classes exist in a flat space. In other words, just as parent concepts may have many child concepts, parent elements may have many child elements. In contrast, base class instance in object-oriented approaches cannot be shared among child class instances (for each parent there is only one child). In this sense CoM resembles the hierarchical data model where elements also exist within a hierarchy. The difference is that the purpose of the CoM inclusion hierarchy consists in describing how elements are identified while data semantics is represented using multi-dimensional structure described below. Inclusion relation allows us to describe a hierarchical address system like conventional postal addresses.

Concepts declare inclusion relation but the collections need to be appropriately *bound* so that the entity segments are really hierarchically connected. This is done during collection creation analogically to concept declaration. For example, if we create a new table with accounts then we have to specify a table where parent elements are stored:

```
CREATE TABLE Accounts CONCEPT Account IN Banks
```

Now if we select accounts and retrieve also some fields from the parent elements (banks) then the parent fields will be taken from table `Banks`:

```
SELECT balance, parent.name FROM Accounts
```

This SQL-like query retrieves all accounts with their balance and their bank name. Keyword 'parent' points to the bank where this account exists.

The dual multi-dimensional structure is described by ordering relation between concepts, i.e., each concept is supposed to have a number *super-concepts* and a number of *sub-concepts*. In order to establish such a structure we use the following main principle: *each field type of a concept specifies one super-concept*. Thus a concept has as many super-concepts as it has fields and it has as many sub-concepts as it is used as a field type in other concepts. For example (Fig. 3b), if banks are characterized by an address then concept `Address` is a super-concept of concept `Bank`:

```
CONCEPT Address IN City                    // Super-concept for Bank
  IDENTITY CHAR(10) addressCode
  ENTITY CHAR(54) street, CHAR(8) house

CONCEPT Bank                               // Sub-concept for Address
  IDENTITY CHAR(10) iban
  ENTITY CHAR(64) name, Address address
```

As a consequence, each concept instance will be an element of the ordered set. At the same time, all elements exist within a hierarchy described by concept inclusion relation. Each item (concept instance) references many super-items using its fields. Here we see one important difference from other approaches: fields and references are interpreted as elements of ordering relation and hence the whole model gets semantics. Other approaches use references to connect elements in an arbitrary graph rather than an ordered structure.

In diagrams, the multi-dimensional structure is drawn as a graph where nodes are concepts while edges are connections from a concept to its super-concept. All edges are drawn as upward arrows and hence more general super-concepts are positioned above their sub- concepts. For example, if concept `Bank` has a filed specifying its address with the type of concept `Address` then this means that `Address` (referenced type) as a super-concept of `Bank` (referencing type). In this case we draw sub-concept `Bank` below super-concept `Address` and connect then by an upward arrow as shown in Fig. 3b. If concept `Person` also specifies concept `Address` as a type of one of its fields then concept `Address` is a super-concept for both `Bank` and `Person`.

Just as concepts, collections exist as elements of an ordered set. When a new collection is being created we need to *bind* it to its super-collections. This can be done by assigning the corresponding fields of the collection when it is being created:

```
CREATE TABLE Accounts CONCEPT Account IN Bank
  owner = Persons                          // Bind to super-collection
```

In this statement we say that elements of this collection will reference elements of collection `Persons` in their field `owner` (and have parents in collection `Banks`).

*Concept schema* is a number of concepts where each concept belongs to two structures: a hierarchy using inclusion relation and an ordered set using field types. Note that concept inclusion hierarchy is precisely what the concept-oriented programming deals with (CoP also uses substitution relation for describing address resolution and virtual addresses). An example of concept schema used in the next sections is shown in Fig. 4. Here we assume that an address exists in the context of some city, i.e., the first segment of any address is city while the second segment is street, house number etc. Persons and banks are characterized by their address. Accounts are defined in the context of their bank and hence concept `Account` is included in concept `Bank`. We assume that there is many-to-many relationship between accounts and their owners and therefore concept `AccountOwner` has two super-concepts: `Person` and `Bank`.

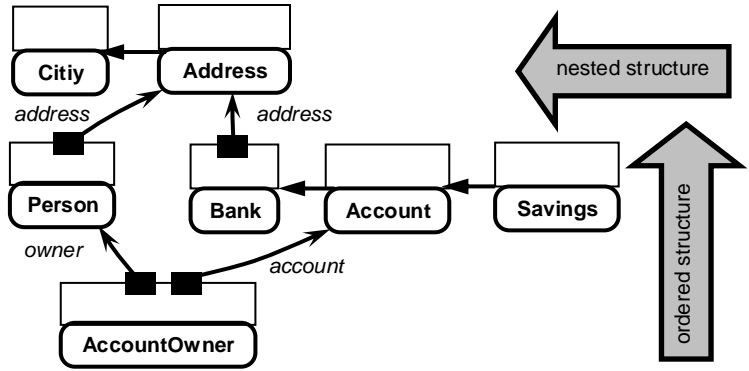

**Fig. 4.** Concept schema.

### 4.2 Logical Navigation

By *physical navigation* we mean access to data using their (complex) identities which determine position of element in the hierarchical structure. One segment is one step in physical access path. This style is used in the hierarchical data model and was shown to be very efficient but rather inflexible. By *logical navigation* [Sav05b] we mean access to data without the need to specify their identities or join conditions but using only the multi-dimensional structure of the model. One dimension is one step in logical access path.

CoM provides a very simple and flexible mechanism for logical navigation which is based on operations of projection and de-projection. The idea is that we can use dimension names as navigation paths for moving over the model structure between different collections. A sequence of projections and de-projections is referred to as (logical) *access path*. One segment in access path is either projection or de-projection operation denoted by '->' and '<-', respectively:

```
SourceCollectoin -> dimension -> SuperCollection
SourceCollectoin <- sub-dimension <- SubCollection
```

Collections in an access path can be the result of some query. The simplest approach consists in restricting elements of an existing collection by imposing some constraint separated by bar symbol:

```
(Collectoin s | SomeCondition(s) == true)
```

Here `s` is an instance variable which might be needed to express the selection criteria.

In order to navigate over hierarchical structure keyword 'parent' can be used as a special dimension connecting child element with its parent element:

```
SourceCollectoin -> parent -> ParentCollection
SourceCollectoin <- parent <- ChildCollection
```

The first query returns all parents of the source collection while the second query returns all children for the elements of the source collection.

Suppose (Fig. 4) we have a collection of addresses in Berlin and want to find all bank accounts related to them for persons older than 20 in banks with address in Bonn. This can be done by de-projecting these addresses down to `AccountOwners` and then projecting up to `Accounts` with the corresponding intermediate restrictions:

```
(Addresses | city = 'Berlin')                          // Source collection
  <- address <- (Persons | age > 20)
```

```
    <- owner <- AccountOwners
    -> account -> (Accounts | parent.address.city = 'Bonn')
```

The result set is a collection with the found account (complex) identities. We can apply aggregation functions to collections and use projection and de-projection to restrict selected items. For example, for the above query we might be interested in finding only accounts with at least two owners and 100 EUR on savings sub-account. In this case the last projection of the above query is modified as follows:

```
    -> account -> (Accounts | parent.address.city = 'Bonn' AND
      this <- account <- AccountOwners > 2 AND
      SUM(this <- parent <- SavingsAccounts.balance) > 100
      )
```

Here de-projection `this <- account <- AccountOwners` returns all account owners and comparison with 2 is a shortcut for `SIZE` (of collection). The last line finds all savings sub-accounts for which we select only one numeric field (balance) and then compute the total balance for them using `SUM` function.

This example shows how rather complex queries can be written in very concise and natural form. Notice that we used many collections but no joins were needed because data is already stored within hierarchical and multi-dimensional structure which is used for access.

### 4.3 Multi-Dimensional Analysis

In many applications such as multi-dimensional analysis and OLAP we need to produce *new* collections as product of existing ones [Sav06a, Sav06b]. In CoQL source collections are written in round brackets separated by comma. Optionally this list can start from keyword FORALL. For example, we could build a new collection as a product of all cities and all banks:

```
    Collection ResultCube = ( Cities, Banks )
```

Each element of a new multi-dimensional collection is treated a cell of a multi-dimensional cube for which we can assign a parameter, called measure in OLAP. The measure can be computed as an additional dimension of the collection:

```
    ( Cities city, Banks bank, measure = <expression> )
```

Here `<expression>` has to return some value (or element) given instance variables `city` and `bank` (which are also field names in the new collection). An equivalent more verbose syntax for this operation breaks a query into several sections:

```
    FORALL ( Cities city, Banks bank )                          // Product
    WHERE ( predicate(city, bank) == true )                     // Criteria
    BODY ( DOUBLE measure = <expression> )                      // Query body
    RETURN ( city, bank, measure )                              // Output
```

FORALL is product of source collections. WHERE provides selection criteria where we can choose what elements of this product to return. RETURN indicates the necessary fields in the new multi-dimensional collection, i.e., what super-items to use. BODY block is used for computing intermediate values or collections for each element in the product.

Suppose (Fig. 4) that we want to build a diagram with cities and banks as horizontal axes. As a measure, this diagram has to draw the total account balance of all persons from this city owning an account in this bank. Note that a person may have accounts in many banks and one account can be owned by many persons. Data for this diagram can be produced using the following query:

```
    FORALL ( Cities city, Banks bank )                          // Dimensions
    BODY (
      Collection CityAccounts =
        city <- parent <- Addresses                             (1)
        <- address <- Persons                                   (2)
        <- owner <- AccountOwners                               (3)
        -> account -> (Accounts | parent.bank == bank )         (4)
      measure = SUM( CityAccounts.balance )                     (5)
      )
    RETURN ( city, bank, measure )
```

Here we start from the `city` (1) and find all its account owners by de-projecting to `Persons` (2) and then to `AccountOwners` (3). Then these account owners are projected to `Accounts` but we select only those in the current bank (4). Finally, we need to aggregate data in this intermediate collection by summing up their balance and storing it in variable `measure` (5) which is returned as the third dimension.

## 5   Related Work

The main difference of CoM from all existing data models is that it relies on duality principle which separates identity and entity modelling using two orthogonal structures. The hierarchical structure of inclusion relation is similar to the hierarchical model of data. In particular, the main purpose of the both consists in describing how elements are represented and accessed.

Another major distinguishing feature of CoM is based on order theory for describing data semantics. The role of ordering relation in data modelling was studied in [Zan84, Bun91] where partial order is a consequence of having incomplete information in data. However, this research was restricted by the frames of the relational model and did not produce new foundations for data modelling. In contrast, CoM assumes that any data is inherently ordered as one of its fundamental properties and then we develop this formal order-theoretic setting into a data model by adding the necessary features and mechanisms.

Order theory is a foundation of formal concept analysis (FCA) [Gan99,Wil06]. However [Sav05a], in FCA formal concepts are derived from data, i.e., the structure of concepts changes if data is modified. In CoM, a fixed ordered structure is defined before it is filled in with data.

The universal relation model (URM) [Ken81, Fag82, Mai84] was aimed at introducing a kind of canonical representation in the form of a universal relation where all relations are viewed as its projections. However, the assumption of universal relation was shown to be incompatible with many aspects of the relational model and did not result in a new foundation for data modelling. CoM also introduces a kind of universal relation but makes it on order-theoretic basis and in this sense it reaches general goals of URM.

The navigational features of CoM make it similar to the functional data model (FDM) [Shi81, Gra99, Gra04]. The difference is that logical navigation in CoM is based on projection and de-projection while navigation in FDM is graph-based in is closer to the network model.

There exist numerous approaches to describing multi-dimensional data most of which rely on the notion of dimension and data cube [Li96, Agr97, Gys97, Ngu00, Tor03, Mal06]. CoM can be viewed as a new approach to multi-dimensional modelling with the main difference is that it is an integrated model (rather than a specific mechanism) based on order theory.

## 6   Conclusions

In this paper we described an approach to data modelling based on nested ordered sets which allows for describing two sides the data world: representation and semantics. We also described simple query language and demonstrated how it can be used to solve some typical problems. Its main advantage is that using a few general notions we can describe many mechanisms and patterns of data modelling. Identity modelling has been described very concisely but this direction is as important as entity modelling and data semantics. Identity modelling in CoM is what concept-oriented programming (CoP) deals with [Sav05c, Sav07a, Sav08, Sav09b]. Both CoM and CoP have been developed within one initiative with the goal to decrease existing differences between programming and data modelling.